\documentclass[amsmath,amssymb,twocolumn,longbibliography,notitlepage]{revtex4-1}
\usepackage{graphicx}
\usepackage{subfigure}

\usepackage{color}
\usepackage{rotating}
\usepackage{physics}
\usepackage{floatrow}
\usepackage{amssymb}
\usepackage{xfrac}

\begin{document}

\title{Constraining domain wall dark matter with a network of superconducting gravimeters and LIGO}

\author{Rees L. McNally}
\email{rm3334@columbia.edu}
\affiliation{Department of Physics, Columbia University, 538 West 120th Street, New York, NY 10027-5255, US}

\author{Tanya Zelevinsky}
\affiliation{Department of Physics, Columbia University, 538 West 120th Street, New York, NY 10027-5255, US}

\begin{abstract}
There is strong astrophysical evidence that dark matter (DM) makes up some 27\% of all mass in the universe.  Yet, beyond gravitational interactions, little is known about its properties or how it may connect to the Standard Model.  Multiple frameworks have been proposed, and precision measurements at low energy have proven useful to help restrict the parameter space for many of these models.  One set of models predicts that DM is a scalar field that ``clumps" into regions of high local density, rather than being uniformly distributed throughout the galaxy. If this DM field couples to a Standard Model field, its interaction with matter can be thought of as changing the effective values of fundamental constants. One generic consequence of time variation of fundamental constants (or their spatial variation as the Earth passes through regions of varying density) is the presence of an anomalous, composition-dependent acceleration.  Here we show how this anomalous acceleration can be measured using superconducting accelerometers, and demonstrate that $>20$ years of archival data from the International Geodynamics and Earth Tide Services (IGETS) network can be utilized to set new bounds on these models. Furthermore, we show how LIGO and other gravitational wave detectors can be used as exquisitely sensitive probes for narrow ranges of the parameter space. While limited to DM models that feature spatial gradients, these two techniques complement the networks of precision measurement devices already in use for direct detection and identification of dark matter.
\end{abstract}

\maketitle

\section*{Introduction}

There is an abundance of cosmological evidence for the existence of dark matter (DM), yet we have been unable to probe its nature in the laboratory. As the search for WIMPs approaches the background limits set by neutrino interactions \cite{OHare2016}, it has become increasingly important to consider a broader range of models and experimental platforms.  One set of models attributes DM to the existence of  an ultralight field with mass ranging from $10^{-22}$ to $10^{4}$ eV.  An example of such a field is the axion \cite{Sikivie1983} which has been the focus of tremendous experimental effort \cite{Graham2016}.  In this work, we consider the case where the ultralight scalar field forms spatially inhomogeneous structures \cite{Battye1999,Sikivie1982,Perelstein2003}, and couples quadratically to a Standard Model field. This results in apparent changes of fundamental constants as the Earth travels through areas of varying field density. These changes would in turn cause anomalous accelerations we can measure in the lab, using superconducting gravimeters (SCG) in the IGETS network, and with gravitational wave detectors such as LIGO and LISA. This search technique complements existing efforts to detect structured ultralight DM using atomic clocks \cite{Roberts2019,Wciso2016,Derevianko2014,Roberts2017} and atomic magnetometers \cite{Banerjee2019,Afach2018, Afach2018,JacksonKimball2018,Pustelny2013,Pospelov2012}. 
We point out (as derived in Eq. (11)) that a large enhancement can be achieved with acceleration measurements for a range of domain wall sizes $d$:
in terms of the fractional change in a fundamental constant, $\epsilon$, the acceleration scales as $\epsilon c^2/d$. While 2D (sheet-like), 1D (line-like), and 0D (point-like) spatial distributions of ultralight scalar fields are possible in stable formations, we focus on the domain wall (2D) distribution \cite{Press1989}.

\section*{Quadratic scalar couplings and fundamental constants}

We focus on fields with quadratic scalar couplings to the Standard Model  because limits on linear scalar coupling are already stringent, and derivative couplings are better suited for spin-dependent searches \cite{Pospelov2012}. Neglecting interactions at lower orders, we can denote the interaction Lagrangian between an ultralight DM field at a specific position and time, $\phi(r,t)$, and any Standard Model (SM) field as
\begin{equation}  -\mathcal{L}^{DM-SM} = \phi^2 (r,t)\bigg( \Gamma_{f}m_{f}c^2 \bar{\psi_{f}}\psi_{f} - \frac{\Gamma_{\alpha}}{4}F_{\mu \nu}F^{\mu \nu} + ... \bigg),\end{equation}
where $m_{f}$ and $\psi_{f}$ are the mass and field for each fermion (with an implied sum over the fermions), $F_{\mu \nu}$ is the electromagnetic tensor, and $\Gamma_{X}$ is the coupling constant between the $X$th component of SM field and the DM field.  Comparing this to the SM Lagrangian
\begin{equation}  -\mathcal{L}^{SM} =  m_{f}c^2 \bar{\psi_{f}}\psi_{f} + \frac{1}{4}F_{\mu \nu}F^{\mu \nu}+...,\end{equation}
we observe that to lowest order this DM-SM interaction acts by changing the effective value of the coupling constant for each field. Stated another way, the presence of the DM field has the effect of shifting the apparent values of fundamental constants in the following way:
\begin{equation} \alpha^{eff} = \alpha \bigg(\frac{1}{1-\Gamma_{\alpha}\phi(r,t)^2}\bigg) \approx \alpha(1 + \Gamma_{\alpha}\phi(r,t)^2), \end{equation}
\begin{equation} m^{eff}_f = m_f(1 + \Gamma_{f}\phi(r,t)^2),\end{equation}
where $m_f$ can refer to the mass of the electron, proton, or neutron (the three terms our measurement will be sensitive to), and $\alpha$ is the fine-structure constant which captures the strength of the electromagnetic interaction.

For clumpy DM models, $\phi(r,t)$ can generate spatially varying effective values for the fundamental constants.  If we assume that this field constitutes all of DM, and that each defect has the same peak field amplitude ($\phi_{max}$), then we can relate the size of the defect to $\phi_{max}$ using the known density of DM over galactic scales. We begin by expressing the energy density inside the DM field defect with a characteristic size $d$ (related to the mass scale of the DM field by its Compton wavelength, $d \sim \hbar/(m_{\phi} c)$), as 
\begin{equation} \rho_{inside} = \frac{\phi_{max}^2}{\hbar c d^2}.\end{equation}
Since the average energy density over many defects must be equal to the overall energy density measured over the Milky Way of  $\rho_{DM}\approx0.4$ GeV/cm$^3$, we can connect the overall DM density to the density inside a defect of dimension $n$ (where $n=0$ is a point defect, $n=1$ is a line defect, and $n=2$ is a domain wall defect) using simple scaling arguments \cite{Derevianko2014},
\begin{equation}  \rho_{DM} \approx \rho_{inside} d^{3-n}L^{n-3}(\hbar c)^{-1} \approx \phi_{max}^2 d^{1-n} L^{n-3} (\hbar c)^{-1},\end{equation}
where L is is the typical separation between defects, which we can relate to the expected time between defect encounters $\tau$ based on the velocity of the Earth relative to the galactic frame, $v_r\approx300$ km/s.  So for a given defect size and geometry we know the amplitude of the field inside the defect. This can be related to the fractional change in fundamental constants inside a defect, where we focus on domain walls ($n=2$): 
\begin{equation} \frac{\Delta m_f}{m_f} = \Gamma_{f}\phi_{max}^2 = d\tau v_r\hbar c\Gamma_f \rho_{\mathrm{DM}},\end{equation}
\begin{equation} \frac{\Delta \alpha}{\alpha} = \Gamma_{\alpha}\phi_{max}^2 = d\tau v_r\hbar c \Gamma_{\alpha} \rho_{\mathrm{DM}}.\end{equation}
This allows us to relate any measured change in fundamental constants to the coupling between $\phi(r,t)$ and the standard model ($\Gamma_f$), and the distribution of the DM defects in the galaxy (defined by $\tau$ and $d$). Every other parameter is known. Furthermore, the length scale of the defect $d$ and the mass of the field are related by the Compton wavelength. 
For comparison to other work, we will express limits in the effective energy scale, which is related to the the coupling constant,
\begin{equation} \Lambda_X = \frac{1}{\sqrt{| \Gamma_X |}}. \end{equation}

\section*{Anomalous Acceleration from Changes in Rest Mass}

The changes in the effective mass of fundamental particles will also change the masses of macroscopic objects. This has a pronounced effect on their motion, because conservation of energy requires a force associated with the gradient of the object's rest mass energy,
\begin{equation}
\vec{a} = \frac{-\vec{\nabla}m c^2}{m}.
\end{equation}
This expression has significant consequences \cite{Uzan2002,Nordtvedt1990} as it implies an apparent violation of the universality of free fall: since $\nabla M$ is not the same for different fermions, the anomalous acceleration is composition-dependent. 

The test mass composition determines how a fractional change in fundamental masses affects the total mass, and a fractional change in the fine structure constant affects the electronic binding energy of each atom in the test mass. This analysis will be presented in Table 1.  For simplicity, here we focus on the overall fractional change in a test mass.  For a defect of size $d$ that provides a fractional change $\epsilon$ of the test particle mass,
the maximum acceleration is
\begin{equation}
\abs{a}  \approx \frac{\epsilon c^2}{d}.
\end{equation}
This means gravimeters have a high sensitivity to these signals for defect sizes when $c^2/d >> g$.  Therefore, this technique can compete with high-precision measurements such as atomic clocks, since the fractional change we are searching for is enhanced relative to $\epsilon$.  To emphasize this point, domain wall dark matter search using GPS satellites \cite{Roberts2017} is most sensitive to a fractional change in fundamental constants on the order of $10^{-12}$ for a defect size of $10^4$ km. These parameters would result in an acceleration of $10^{-2}$ m/s$^2$ that lasts $\sim30$ s.  This is a large acceleration, for quite a long duration. For comparison, the SCGs that make up the IGETS network are capable of acceleration measurements of $10^{-11}$ m/s$^2$ over minute timescales \cite{Goodkind2012,Virtanen2018}, making them very competitive for these searches using existing technology and data sets.  We now perform a more careful analysis for this measurement.

\section*{Accelerometer Based Fractional Mass Limits}

A natural platform to search for this anomalous acceleration is an accelerometer network.  For the sensitivity estimates, we assume the density of a domain wall follows a Gaussian distribution, with length scale $d$ and a maximum fractional change in the rest mass of $\epsilon$. This choice of distribution can be modified but, as long as the density is relatively smooth, the results will not be strongly affected.  For a test particle traveling perpendicularly through such a domain wall at speed $v_r$, we can define the effective mass as a function of time,
\begin{equation}m_{\mathrm{eff}}(t) = m_0(1 + \epsilon e^{-t^2 v_r^2/d^2}),\end{equation}
where $m_0$ is the unperturbed mass.
Then we calculate the effective acceleration of this test particle as a function of time using Eq. (11), keeping only the leading order in $\epsilon$,
\begin{equation}a(t) = \frac{2\epsilon t v_r c^2}{d^2} e^{-t^2v_r^2/d^2}.\end{equation}
The fractional change in test mass (Eq. (13)) and the effective acceleration waveform (Eq. (14)), as well as a cartoon of a passing domain wall, are shown in Fig. 1.

\begin{figure}[h]
\centering
\includegraphics[width=1 \textwidth]{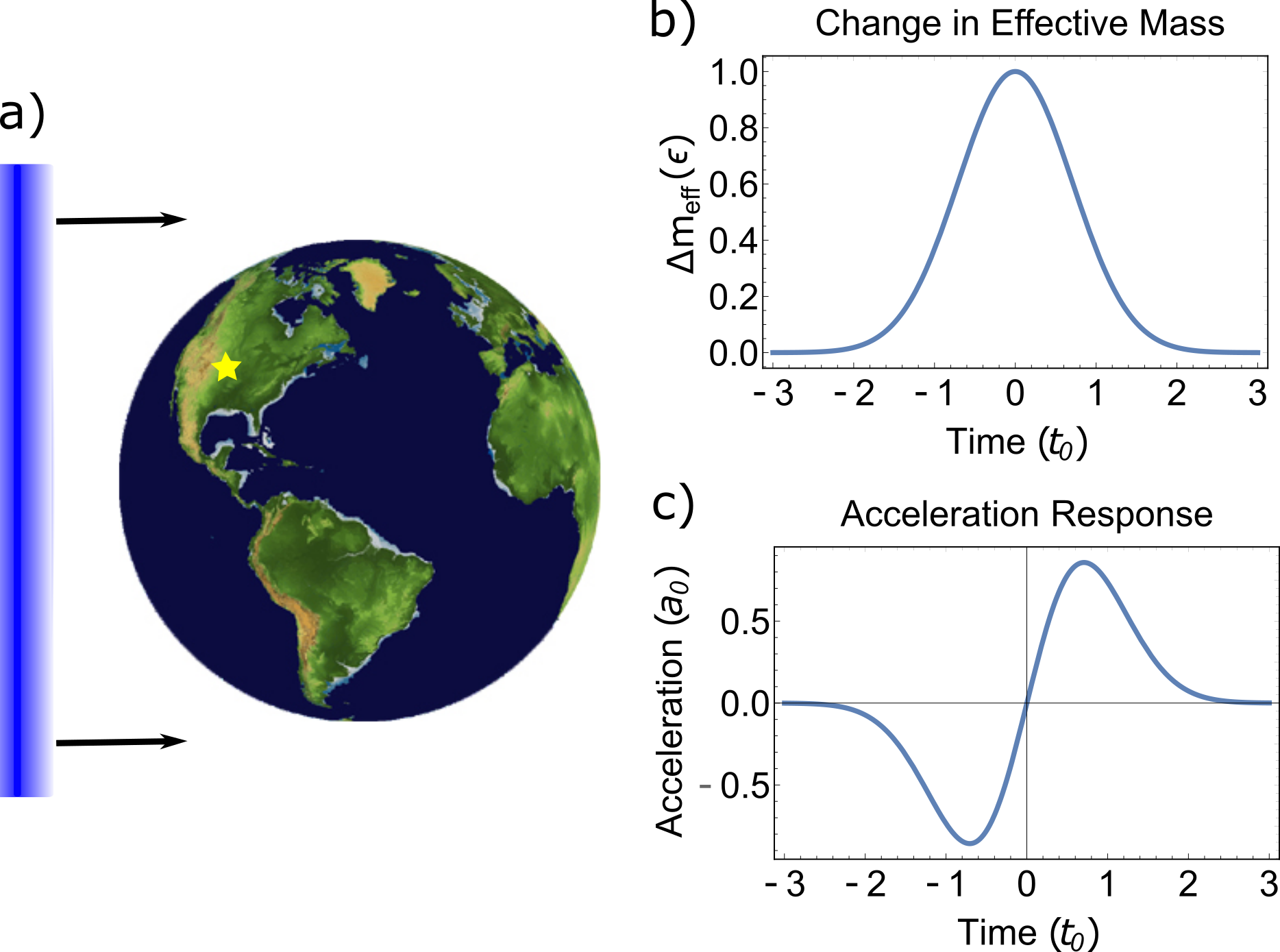}
\caption{a) A domain wall approaching the Earth, with a highlighted accelerometer located in Boulder, Colorado, USA.  b) Fractional change in test mass caused by the domain wall's passage ($\Delta m_{\mathrm{eff}}/m_0 = m_{\mathrm{eff}}/m_0 - 1$).  c) Anomalous acceleration caused by the gradient in the domain wall density, resulting in a changing mass of the test particle. The transient signal has a characteristic time and acceleration of $t_0 =d/v_r$ and $a_0 = \epsilon c^2/d$, respectively, with a peak acceleration of $a_{\mathrm{peak}} =a_0\sqrt{\frac{2}{e}}$ because of the specific density distribution chosen.}
\label{fig:close}
\end{figure}
To detect the anomalous accelerations,
we propose using superconducting gravimeters (SCGs), thanks to their low-frequency stability and extremely high precision. They have recently been used for several tests of local Lorentz violation \cite{Flowers2017,Shao2018}, and in searches for other forms of dark matter \cite{Horowitz2019,Voigt2019}.  These sensors work by monitoring the location of a superconducting niobium sphere relative to a set of aluminum pickup coils \cite{Goodkind2012}.  Because the niobium sphere is free to move while the pickup coils are fixed to a solid structure, a passing domain wall would accelerate the sphere but not the rest of the device, and the sensor would pick up this acceleration. 

For the quantitative estimates of sensitivity, we take the noise spectra for a SCG in the IGETS network during a quiet 10-day period after known tidal forces and local disturbances were removed \cite{Banka}.  This sensor is typical of the $\sim20$ SCGs that comprise the IGETS network, which has been collecting data continuously for the past two decades.  In order to be observed, the domain wall must produce a signal with high enough power in the frequency range where the accelerometer is sensitive. However, the frequency components of this waveform are quite broadband.  This means that for large domain walls the signal will have more power at low frequencies, and for a specific domain size it will be well matched to the $\sim0.01$ Hz peak sensitivity of the SCGs.  To calculate the sensitivity based on the frequency performance of these sensors, we use the waveform in Eq. (14) and take its Fourier transform. We then find the signal-to-noise ratio (SNR) at each frequency by dividing this signal amplitude by the noise amplitude, and calculate the total SNR by integrating over all frequencies.  Finally, we multiply by $\sqrt{t_0}$ to account for the characteristic time over which the signal can be averaged, because the signal we are looking for is transient. We then solve for the value of $\epsilon$ that would result in a signal with an SNR of 10 for a fixed defect size.

Note that for a specifically designed acceleration based experiment, we could expect a sensitivity improvement of several orders of magnitude.  This is because the SCGs do not use the free-fall-universality violating nature of the signal to cancel disturbances.  In particular, a torsion balance experiment with a test mass consisting of two different materials would be a natural setup for these searches, similar to previous axion searches \cite{Terrano2019}.  However, such a system would have to run continuously for several years before it is sensitive enough to limit rare domain wall events. 

\section*{LIGO Based Fractional Mass Limits}

\begin{figure}[h]
\centering
\includegraphics[width=1 \textwidth]{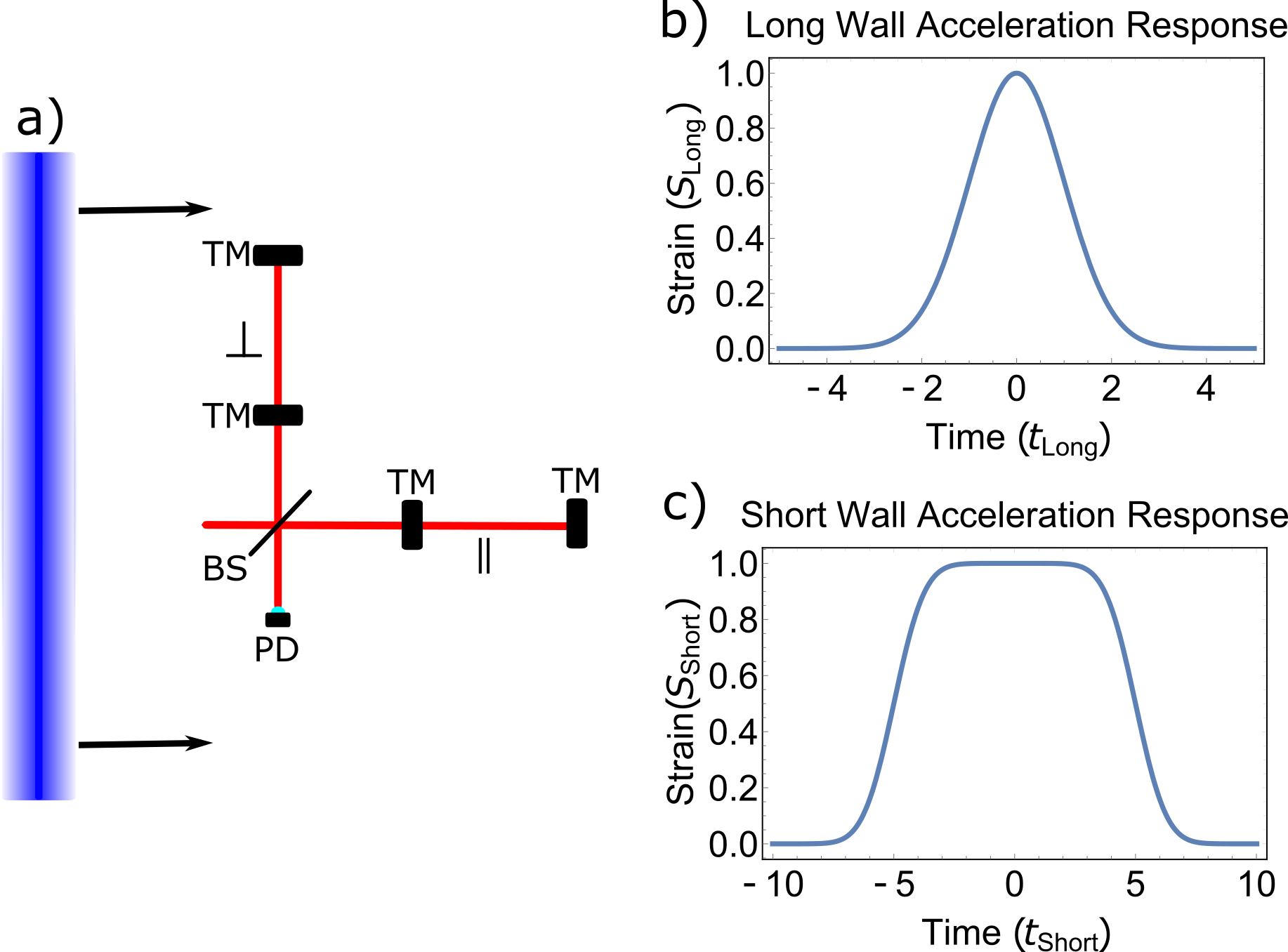}
\caption{(a) Domain wall approaching a LIGO type optical interferometer, along one of the directions of maximum sensitivity. The interferometer is formed by splitting laser light with a beam splitter (BS), while the ends of each arm are free floating mirrors acting as test masses (TMs). Differential changes in the length of each arm are monitored via a photodiode (PD).  In the case shown, the test masses along the arm perpendicular to the domain wall's approach direction ($\bot$) will feel no acceleration along the direction of the arm, while the two test masses in the parallel arm ($\parallel$) will feel slightly different accelerations because of the difference in the gradient of the domain wall at each location. This results in a differential acceleration of the TMs in the parallel arm which changes the path length, while the perpendicular path length is unaffected.  This change in relative path length would be observed as a transient strain in the detector as the domain wall passes.  The resulting (normalized) strain wave forms show qualitatively different behavior when the domain wall is larger (b) or smaller (c) then the path length of LIGO ($L$ = 4 km).  For long domain walls the signal approaches a Gaussian with a peak strain size of $S_{\mathrm{Long}}= \epsilon c^2/v_r^2$. For short domain walls the plateau has a fixed duration $t_p=L/v_r$, and the peak strain scales like $S_{\mathrm{Short}} = \epsilon d c^2/(L v_r^2)$. For both signals, the characteristic time is $t_{\mathrm{Short}} = t_{\mathrm{Long}}= d/v_r$. For wave forms shown, $d\gg L$ for the long domain wall and $d \approx L/10$ for the short domain wall. }
\label{fig:close}
\end{figure}

Another strategy for detecting anomalous accelerations is to use LIGO or other gravitational wave detectors. This has been previously analyzed based on comparisons of the interferometer phase relative to a local oscillator \cite{stadnik2015searching,stadnik2016enhanced}, and was recently proposed based on a similar approach to this paper \cite{Grote2019}.  For example, suppose a domain wall approaches the LIGO detector, perpendicular to one arm of the interferometer and parallel to the other, as shown in Fig. 2 . For the arm perpendicular to the domain wall's propagation direction, each mirror will accelerate by the same amount but the length of the arm will not change.  For the arm parallel to the domain wall propagation, the mirrors will see slightly different field magnitudes (because of the 4 km separation) and therefore experience different accelerations.  At $t=0$ when the domain wall is centered between the two mirrors, the differential acceleration is
\begin{equation}
\Delta a(t) =  a(t - L/2v_r) - a(t + L/2v_r).
\end{equation}
This differential acceleration can be integrated twice to find the change in separation between the mirrors.  Divided by the length of the interferometer $L$ it is the strain $S(t)$,
\begin{equation}
S(t) =  \frac{\epsilon d \sqrt{\pi} c^2}{2 L v_r^2}\bigg[\erf\bigg(\frac{L - 2 t v_r}{2 d}\bigg)+\erf\bigg(\frac{L + 2 t v_r}{2 d}\bigg)\bigg],
\end{equation}
where $\erf(x)$ is the error function. This differential strain between the two LIGO arms is exactly what it was designed to measure extremely precisely.  The strain signal in the detector has qualitatively different behaviors depending on whether the defect is larger or smaller than the 4 km length of the LIGO arms as shown in Fig. 2.  We use this strain in Eq. (16) to perform a sensitivity analysis in analogy to what was done for accelerometers, using a power law approximation of LIGO's strain noise curve \cite{Abbott}, and mandating a SNR of 10.  Furthermore, this analysis was repeated for the planned LISA mission based on its projected sensitivity.

\section*{Atomic Rest Mass and Fundamental Constants}

Before we derive the constraints based on the described technique, we must estimate how a fractional change in each fundamental constant affects the total test mass.  This effect is composition-dependent.  For LIGO the test masses are the silica mirrors \cite{Aasi2015}, for LISA they are Au-Pt composites\cite{Grote2019}, and for the accelerometers they are niobium (the test mass that is  being levitated)\cite{Goodkind2012}.  The semi-empirical Bethe-Weizacker formula \cite{Uzan2002} accounts for the nucleon mass, the electron mass, and the mass associated with the binding energy,
\begin{equation}
\begin{split}m(A,Z) \approx Zm_p +  (A-Z)m_n + \\ Zm_e +  (98.25\;\mathrm{MeV}/c^2)\frac{Z(Z-1)}{A^{1/3}}\alpha,\end{split} \end{equation}
where A is the total nucleon number and Z is the number of protons.  This allows us to calculate the sensitivity coefficient for the $X$th coupling to the DM field, $A_X$, to relate a change in a fundamental constant to the changes in test mass,
\begin{equation}
\epsilon = A_X \frac{\delta X}{X}.
\end{equation}
The results of this calculation for each system are shown in Table 1.

\begin{table}[h]
\centering
\begin{tabular}{|l|l|l|l|}
\hline
Constant              		& LIGO                  & LISA                  & SCG                   \\ \hline
$A_{\alpha}$		        & $5\times10^{-4} $     & $2\times10^{-4} $ 	& $4\times10^{-4} $  	\\ \hline
$A_{m_e}$             		& $2\times10^{-4} $  	& $1\times10^{-4} $  	& $8\times10^{-5}$  	\\ \hline
$A_{m_p}$                 	& 0.5                   & 0.4                   & 0.4                   \\ \hline
$A_{m_n}$                 	& 0.5                   & 0.6                   & 0.6                   \\ \hline
\end{tabular}
\caption{Sensitivity coefficients quantifying how a fractional change in each fundamental constant relates to the fractional change in the test mass for each experimental platform.  As expected, fractional changes in proton and neutron masses result in order unity changes in the total test mass, while changes in electron mass and $\alpha$ result in smaller changes.}
\end{table}

\section*{Projected Limits on New Physics}

\begin{figure*}[h]
\centering
\includegraphics[width=.9\textwidth]{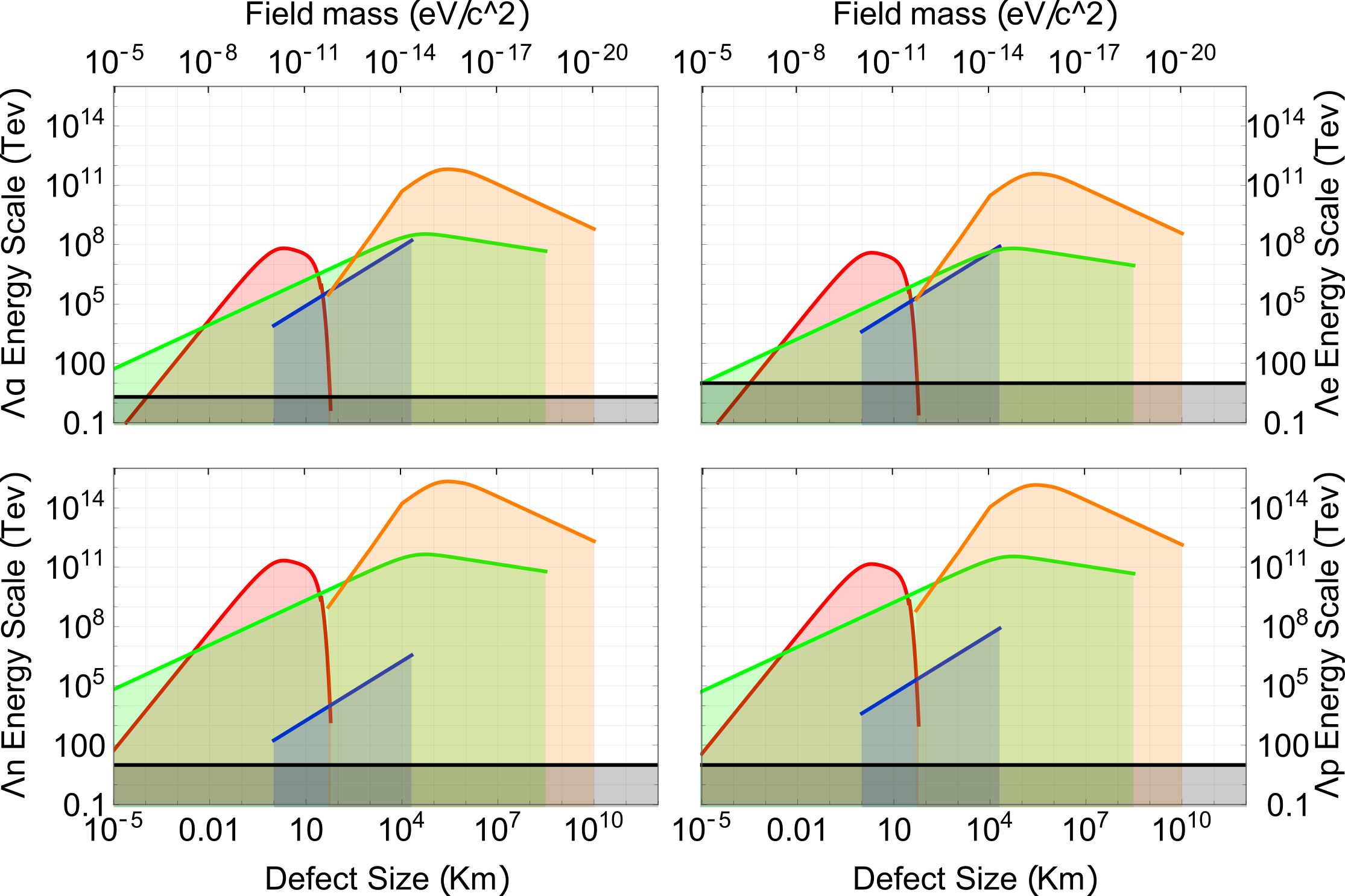}
\caption{Limits on the DM-SM coupling strengths set using GPS clocks \cite{Roberts2017} (blue) and astrophysical constraints \cite{Raffelt1999} (black), and the projected limits set by the IGETS network (green), LIGO (red) and LISA (orange).  Here, the time $\tau$ between domain walls is 7 years to allow a comparison to GPS limits.  We have access to sufficient data taken with the SC gravimeters to look for such rare events, but for LIGO, only $\tau\lesssim1$ year is currently achievable.  We see smaller sensitivities to the $\alpha$ and $m_e$ couplings as they contribute less to the total system mass, making them well suited for clock based searches.  Note that the constraints far from the peak of each system's sensitivity may be unreliable due to additional noise sources not present in the approximated noise power spectral density, or for small defect sizes due to finite sampling rates.}
\label{fig:close}
\end{figure*}
The described analysis allows us to estimate what domain wall size and effective coupling energy scale each experimental platform is sensitive to for the neutron, proton and electron masses and the fine structure constant.  The results are shown in Fig (3).  For these projected constraints we assume that the DM-SM coupling contribution comes entirely from one term, and that the time between domain wall crossings is $\tau=7$  years.  This timing is well matched with the 20 years of IGETS data, and the same was chosen for the previous GPS based studies \cite{Roberts2017}.  Recent work with networks of atomic clocks has also set stringent limits for these coupling models \cite{Roberts2019,Wciso2016}.  However, the limited duration of the data collection period with this network of $\lesssim40$ days (compared to 20+ years of IGETS data) limit their sensitivity to only relatively frequently occurring domain walls.  As more data is collected using this clock network, this situation will improve and will likely be the most accurate way to search for coupling to $\alpha$.

\section*{Conclusions}

We have shown that in addition to atomic clocks and atomic magnetometers, accelerometers and gravitational wave detectors are natural platforms to search for domain-wall ultralight scalar dark matter.  Furthermore, a detection event that is seen in multiple systems simultaneously would provide a strong evidence of interactions beyond the Standard Model.  For the SCG accelerometers, 20+ years of data taken with over 20 sensors has already been recorded and archived.  This will allow us to mine for these signals and investigate an unexplored parameter space for the DM-SM coupling models.  For LIGO, $\gtrsim1$ year of data has been collected, and is also freely available for these searches. This represents a unique opportunity to explore the origins of DM with existing data sets.\\

\section*{Acknowledgments}
The authors are grateful to Ken Van Tilburg for helpful insights into this work, and to Derek Kimball for comments on the manuscript.  R.L.M. gratefully acknowledges support by the NSF IGERT Grant No. DGE-1069240.

\section*{Author contributions}
Both authors contributed to the writing of the manuscript.  R.L.M. performed the signal analysis.





\end{document}